\documentclass[11pt,a4paper]{article}

\usepackage[margin=1in]{geometry}
\usepackage{graphicx}
\usepackage{amsmath,amssymb}
\usepackage{bm}
\usepackage{booktabs}
\usepackage{url}
\usepackage[numbers,sort&compress]{natbib}
\usepackage[colorlinks=true,linkcolor=blue,citecolor=blue,urlcolor=blue]{hyperref}
\usepackage{authblk}
\usepackage{caption}
\usepackage{xspace}

\title{Phase-Separated Complex Hilbert PCA on Markerless 3D Pose Estimation Data:\\
A Global Phase Network and Its Extension to a Continuous Field on the Body Surface}

\author[1]{Hiromitsu Goto\thanks{Corresponding author. \texttt{goto@kanazawa-gu.ac.jp}}}
\author[1]{Tao Tao}
\author[1]{Zheng-Lin Chia}
\affil[1]{Faculty of Information Engineering, Kanazawa Gakuin University, Kanazawa, Japan}

\date{\today}

\begin{document}
\maketitle

\begin{abstract}
Quantitative analysis of the kinematic chain in sports motion is essential for
performance evaluation and injury prevention.  Conventional methods such as
the kinematic-sequence method (KS) and continuous relative phase (CRP) are
confined to anatomically adjacent joint pairs and do not provide a single
framework for describing whole-body coordination, while segmental power-flow
analysis requires force plates and inertial parameters that restrict its use
to laboratory environments.  We apply Complex Hilbert Principal Component
Analysis (CHPCA) separately to each motion phase (backswing and downswing) on
markerless 3D pose estimation data, extracting the dominant whole-body phase
pattern as a single complex eigenvector.  We further implement a fully
automatic signal-based phase segmentation requiring no priors on strike count
or rest location, and extend the analysis to 1{,}079 body-surface mesh
vertices so that the kinematic chain is represented as a continuous phase
field across the body.  Applied to 14 hammer-striking trials of a single
subject, the framework reveals (i) a trunk-anchored global phase architecture
represented as a single complex eigenvector, (ii) a functional asymmetry
between preparation and execution phases quantified by Mode-1 contribution
(45.5\% vs.\ 70.5\%) and inter-trial Spearman consistency (0.38 vs.\ 0.58),
and (iii) a consistent reorganisation across both skeletal joints and mesh
vertices ($p < 10^{-10}$ on 1{,}079 vertices).  As a methodological
consistency check, pairwise phase differences from the Mode-1 eigenvector are
compared against CRP on all 190 joint pairs by a permutation test
($\rho = 0.473$, $p = 0.0005$).  A correspondence analysis between Mode-1
amplitude and kinetic-energy mobilisation variance further shows a strong
positive correlation in the downswing ($\rho \approx 0.71$ on both skeleton
and mesh) and no correlation in the backswing, indicating that the proposed
framework bridges kinematic and kinetic descriptions of coordination through
phase structure.
\end{abstract}

\noindent\textbf{Keywords:}
complex Hilbert principal component analysis,
kinematic chain,
markerless motion capture,
body surface mesh,
phase analysis,
automatic phase segmentation

\vspace{1em}
\noindent\textbf{Note:} A condensed Japanese version of this paper is to be
submitted to \emph{Transactions of the Japanese Society for Artificial
Intelligence} (Special Issue on ``Emerging Topics in Sports Informatics'').
The present arXiv version contains the full analysis and extended discussion;
this page will be updated with a DOI and volume information once the
Japanese version is published.

\section{Introduction}

Quantitatively characterising how motion propagates through body segments---
the kinematic chain, also referred to as the kinetic chain---is central to
both performance enhancement and injury prevention in sports
analysis~\citep{kibler1998-scapula-kinetic-chain,seroyer2010-kinetic-chain-overhead}.

\subsection{Lineage of kinematic-chain metrics}\label{sec:intro_metrics}

Quantification of the kinematic chain can be organised into five families
according to the type of information they capture.  (i) The peak
angular-velocity sequence method (KS) examines the order of peak-velocity
arrivals along an anatomical
chain~\citep{cheetham2008-kinematic-sequence-golf,serrien2018-proximal-distal-meta}.
(ii) Continuous relative phase
(CRP) computes instantaneous phases via the Hilbert transform and describes
the pairwise phase relation between two
joints~\citep{varlet2011-crp-computation,lamb2014-continuous-relative-phase}.
(iii) Vector coding (VC) classifies coordination patterns from the gradients
of joint-angle trajectories~\citep{stock2018-vector-coding,zehr2018-crp-vs-vc}.
(iv) Mechanical-energy and power-transfer
frameworks~\citep{robertson1980-segmental-power,putnam1993-sequential-motion}
quantify the generation, absorption and transfer of energy between segments
in watts.
(v) High-dimensional principal-component approaches extract common modes
across $N$ whole-body variables via
eigendecomposition~\citep{dotov2025-high-dimensional-coordination,kimura2021-coordination-concept}.
These families provide complementary information; however, (i) lacks a
standardised definition of angular velocity
components~\citep{brown2019-angular-velocity-component} and rests on the
questionable assumption that a single optimal sequence
exists~\citep{glazier2019-optimal-technique}, and (ii) and (iii) are
restricted to pairwise analyses.  As exemplified by energy-flow analyses of
baseball pitching across the lower extremities, trunk and throwing
arm~\citep{deswart2022-energy-flow-pitching,aguinaldo2019-segmental-power-baseball},
(iv) requires force plates and inertial parameters and is therefore confined
to laboratory environments.  The classical PCA in (v) captures only
isochronal correlations without temporal precedence or lag between joints.
A method that expresses the phase order of all joints through a single
global description has not been established.

\subsection{Advances of kinematic-chain research through computer vision}\label{sec:intro_cv}

Recent progress in computer vision (CV) has changed the operational
preconditions of kinematic-chain analysis.  Markerless motion capture from
monocular or multi-view video has become
practical~\citep{nakano2020-openpose-markerless,javerliat2025-kineo}, with
joint-centre errors approaching $1$--$3\,\mathrm{cm}$ relative to
optical-marker systems.  Parametric human models such as
SMPL/SMPL-X~\citep{loper2015-smpl} output dense body-surface meshes of
$1{,}079$ vertices or more, beyond the $20$ skeletal joints.  Downstream
applications extend beyond pose recovery to force-plate-free kinetics
estimation---recovering ground reaction forces and joint torques directly
from
video---\citep{katsu2024-grf-monocular-video,liu2024-imdy,pbl2025-portable-biomechanics}
and to differentiable-physics biomechanics that solves pose estimation and
inverse dynamics as a single end-to-end optimisation
problem~\citep{larp2024-neural-physics-pose,cotton2024-differentiable-biomechanics,liu2024-hdys-dynamics-space}.
These developments focus on the measurement side, while the downstream
question of how to describe such dense data in the language of the kinematic
chain is less developed.  Of the five metric families above, few scale
naturally to the $1{,}079$-vertex meshes: KS and CRP depend on fixed joint
definitions, and energy-based methods require per-segment mass and inertia
tensors that are not defined at the vertex level.

\subsection{Position and contributions of this work}\label{sec:intro_contributions}

We apply Complex Hilbert Principal Component Analysis
(CHPCA)~\citep{iyetomi2020-chpca-macroeconomic} to markerless pose
estimation outputs.  CHPCA forms analytic signals via the Hilbert transform
of real time series and extracts the phase relations among many variables
simultaneously through the eigendecomposition of a complex correlation
matrix.  The only required input is the per-point speed-norm time series;
ground reaction forces, segmental masses and rigid-link models are not
needed.  This enables (a) identification of whole-body phase structure not
restricted to bone-adjacent pairs, and (b) a natural extension to all
$1{,}079$ body-surface vertices.  Because the speed norm
$s_i(t) = \|\bm{v}_i(t)\|$ equals the square root of the mass-normalised
kinetic energy ($KE_i/m_i = s_i^2/2$), CHPCA can also be viewed as a
phase-domain description of kinetic-energy mobilisation across the body,
without requiring mass information.

The contributions of this study are:
\begin{enumerate}
\item Quantification of functional asymmetry in coordination structure via
      phase-separated CHPCA (Mode-1 contribution and inter-trial consistency).
\item A framework that describes the kinematic chain as a global phase
      network represented by a single complex eigenvector, lifting the
      pairwise restriction of conventional methods.
\item Extension to a continuous phase field on the body-surface mesh, which
      is practically intractable for CRP due to combinatorial explosion.
\item Empirical verification of the correspondence between CHPCA Mode-1
      amplitude and the variability of kinetic-energy mobilisation, positioning
      the method as a bridge between kinematics-only analysis and classical
      kinetics analysis.
\end{enumerate}
A signal-processing-based phase segmentation that requires no priors on
strike count or rest location is also implemented as a practical component.

\section{Methods}

\subsection{Data}

We used the public demo data of the markerless 3D pose estimation system
Kineo~\citep{javerliat2025-kineo}.  The subject performs a repetitive
two-handed hammer-striking motion that mimics stonework (single subject,
right-handed and right-side striking, in an upright two-footed stance).  The
data consist of $900$ frames ($30\,\mathrm{fps}$, $30\,\mathrm{s}$) of 3D
time-series coordinates in the SMPL-X format ($1{,}079$ keypoints).
Throughout the paper, ``right'' and ``left'' refer to the subject's own body
sides.  For the skeleton-based analysis we extracted $20$ major joints
(pelvis, spine, neck, head, left/right shoulder, elbow, wrist, hip, knee,
ankle and foot).

\subsection{Preprocessing}

To remove translational drift of the positional coordinates we computed
frame-difference velocity vectors $\bm{v}_i(t) = \bm{x}_i(t+1) - \bm{x}_i(t)$
and used the per-point speed norm $s_i(t) = \|\bm{v}_i(t)\|$ as the analysis
input.  The speed norm is a non-directional scalar that yields a single time
series per point.

\subsection{Automatic phase segmentation}\label{sec:autoseg}

To separate each trial of the repetitive motion into backswing and downswing
phases, we implemented a signal-processing-based segmentation pipeline that
uses the $Z$ (height) coordinate of the striking-side wrist as the primary
signal.  The pipeline requires no priors on the number of strikes, rest
location, or dominant hand:
\begin{enumerate}
\item Use \texttt{find\_peaks} with a loose threshold to extract candidate
      maxima (top positions) of the wrist $Z$ trajectory.
\item Apply an elbow-based split on the prominence distribution to separate
      strong tops (true strikes) from weak peaks (noise).
\item A secondary filter discards candidates whose $Z$ height lies in the
      lowest $40\%$ of the signal range.
\item Use $1.8\times$ the median of detected-top intervals as a threshold to
      identify long gaps as rest periods automatically.
\item For each top, the local $Z$ minimum immediately before is labelled as
      the start of the backswing, and the minimum immediately after as the
      impact point ending the downswing.
\end{enumerate}

\subsection{Phase-separated CHPCA}

For each automatically segmented phase (backswing and downswing) we applied
CHPCA using two complementary approaches.  Approach A (concatenated)
concatenates all trials in the same phase along the time axis and performs a
single CHPCA, enlarging the sample size and stabilising the estimate.
Approach B (per-trial) runs CHPCA independently on each trial and aligns the
eigenvector phase to a reference joint (pelvis) before circular averaging,
enabling the quantification of inter-trial variability.

For each segment, both approaches apply the following procedure to the speed
time series $s_i(t)$~\citep{iyetomi2020-chpca-macroeconomic}:
\begin{enumerate}
\item Compute the analytic signal
      $z_i(t) = s_i(t) + j\,\mathcal{H}[s_i](t)$ via the Hilbert transform.
\item Standardise each variable (mean $0$, standard deviation $1$).
\item Form the complex correlation matrix $C = \frac{1}{T} Z^H Z$ ($Z$:
      analytic signal matrix, $H$: Hermitian transpose).
\item Perform the eigendecomposition $C\bm{u}_k = \lambda_k \bm{u}_k$.
\end{enumerate}

From the complex eigenvector of the first mode $\bm{u}_1$ we compute the
Hodge potential $\phi_i = \arg(u_{1,i})$ and amplitude $A_i = |u_{1,i}|$.
The pairwise phase differences $\Delta\phi_{ij} = \phi_i - \phi_j$ between
any two variables are exactly representable as differences of nodal values
$\{\phi_i\}$, and therefore correspond to the gradient (curl-free) component
of the discrete Helmholtz--Hodge decomposition on the variable network;
this is the sense in which $\phi_i$ is referred to as the Hodge
potential~\citep{iyetomi2020-chpca-macroeconomic}.
In Approach B we rotate the eigenvector of each trial so that the reference
joint $r$ (pelvis) has zero phase,
$\bm{u}_1 \leftarrow \bm{u}_1 \cdot e^{-j\arg(u_{1,r})}$, and then take the
circular mean across trials joint-wise.  Inter-trial stability is evaluated
by two indicators.  The first is the per-joint circular resultant length
$R_i = \left| \frac{1}{N_\mathrm{trial}}\sum_{n} e^{j \phi_i^{(n)}} \right|$.
The second is the average pairwise Spearman rank correlation of the full
Hodge-potential vectors between trials, providing a single scalar that
summarises the reproducibility of the lead--lag ordering within a phase.  We
report the latter as ``inter-trial consistency'' in Table~\ref{tab:skeleton}.

Statistical significance of the modes is assessed by rotational random
shuffling (RRS)~\citep{iyetomi2020-chpca-macroeconomic}.  We generate
$n_\mathrm{RRS} = 1{,}000$ null realisations in which each variable's
analytic signal is independently circularly shifted, thereby destroying
inter-variable phase relations while preserving each variable's spectrum.
Real eigenvalues exceeding the $99$th percentile of the null eigenvalue
distribution are deemed significant modes.

\subsection{Network representation of the kinematic chain}

We visualise the CHPCA Mode-1 results as a global kinematic-chain network.
Twenty joints are placed on a canonical pose (the across-trial mean of the
phase-start poses), and node colour encodes the within-phase rank-normalised
$\phi_i$ ($-1$: most lagging, $+1$: most leading).  Bone edges are coloured
by $|\phi_i - \phi_j|$ and drawn as arrows pointing from the leading joint
to the lagging one.

Because CHPCA also provides phase relations for non-bone pairs, we overlay as
dashed arrows the top-$K$ non-bone pairs ranked by the composite score
$s_{ij} = \sqrt{A_i A_j} \cdot |\sin((\phi_i - \phi_j)/2)|$.  This score is
the product of the geometric-mean participation $\sqrt{A_i A_j}$ of the two
variables in Mode~1 and the chord length $|\sin((\phi_i - \phi_j)/2)|$ of
the phase gap, so pairs with near-zero phase difference are automatically
excluded, highlighting long-range pairs that have both high participation and
a well-separated phase.  This is a visualisation heuristic rather than a
standardised indicator, and is used here in an exploratory sense.  Individual
dashed arrows should not be interpreted as ``statistically significant
long-range lead--lag relationships''; the purpose of this subsection is to
demonstrate that a global phase structure not confined to bone adjacency can
be jointly described by a single mode.

The same representation is extended to a continuous heatmap on the $1{,}079$
body-surface mesh vertices, allowing the three resolutions (adjacent CRP,
global CHPCA, continuous-field CHPCA) to be compared side by side.

\section{Results}

\subsection{Validation of automatic phase segmentation}

\begin{figure}[t]
\centering
\includegraphics[width=0.92\textwidth]{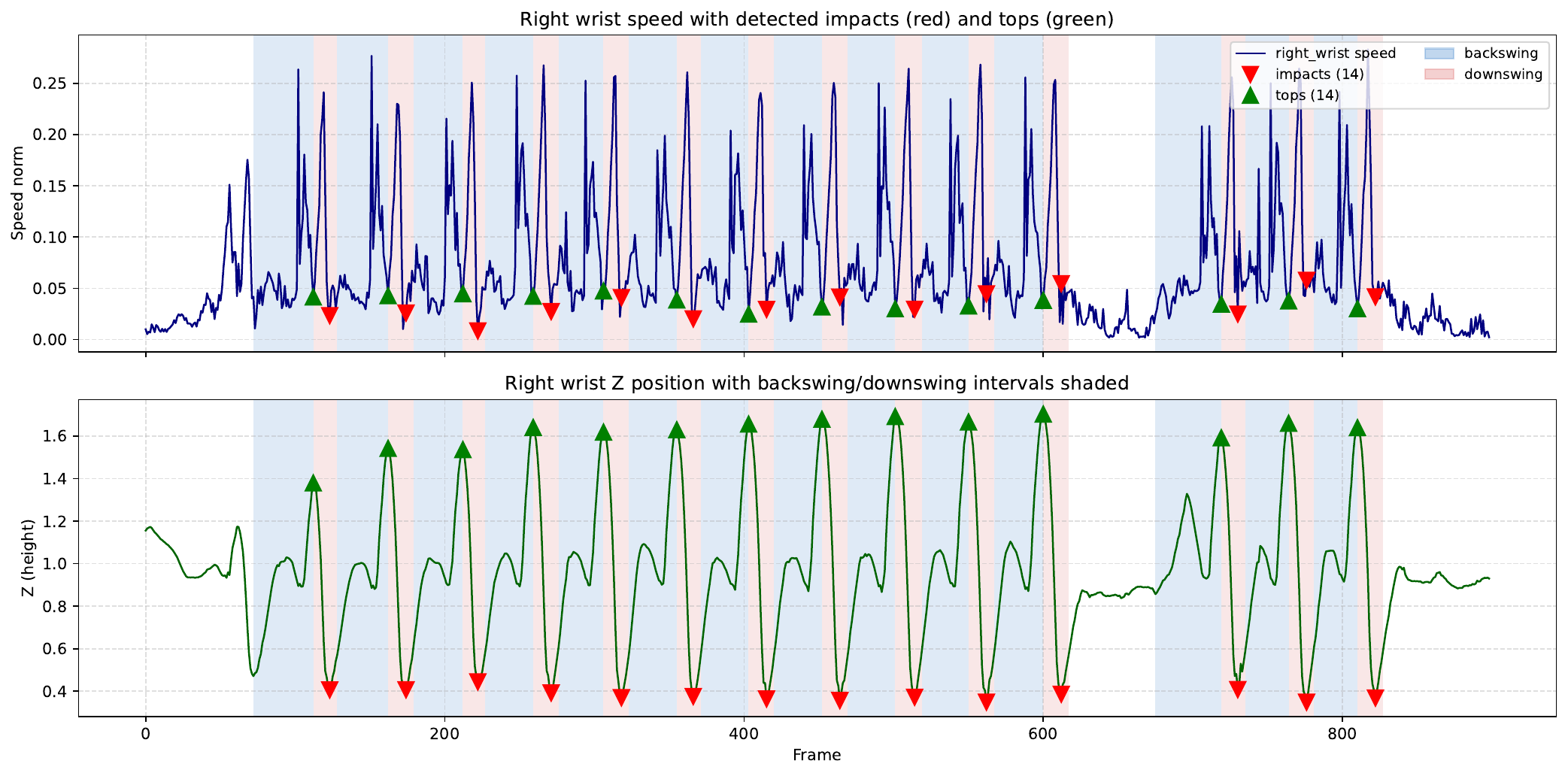}
\caption{Diagnostic plot for the automatic phase segmentation.  Top: speed
norm of the right wrist.  Bottom: $Z$ coordinate (height) of the right
wrist.  Green triangles mark the detected tops and red triangles the strike
points; blue and red shaded bands indicate the backswing and downswing
intervals, respectively.  Without any prior on the number of strikes or the
rest location, $14$ strikes and one rest interval are correctly extracted.
Downswing intervals (red) are nearly constant in duration across trials,
whereas backswing intervals (blue) vary, with trial~12 (the first strike
after the rest) noticeably longer than the others.}
\label{fig:phase_detection}
\end{figure}

Applied to the $900$-frame hammer-striking sequence, the pipeline
automatically detected $14$ valid strikes and one rest interval (frames
$601$--$718$, $\approx 3.9\,\mathrm{s}$; Figure~\ref{fig:phase_detection}).
The detected strikes were automatically grouped into $11$ early-phase and $3$
late-phase strikes.  The per-trial frame ranges and durations of the two
phases are listed in Table~\ref{tab:phase_durations}.  Across the $14$
trials, the backswing duration was $33.1 \pm 4.1$ frames
(range $29$--$44$, $\approx 0.97$--$1.47\,\mathrm{s}$ at $30$\,fps), whereas
the downswing duration was $16.6 \pm 0.7$ frames
(range $15$--$18$, $\approx 0.50$--$0.60\,\mathrm{s}$).  The downswing is
thus not only shorter on average but also far more reproducible in duration
than the backswing; the only outlier in backswing duration is trial~12 ($44$
frames), which immediately follows the rest interval.

\begin{table}[t]
\centering
\small
\caption{Per-trial frame ranges and durations of the automatically extracted
backswing and downswing phases.  Durations are reported in frames; at
$30$\,fps, $30$ frames $\approx 1.0\,\mathrm{s}$.  Trials $1$--$11$ form the
early cluster and trials $12$--$14$ form the late cluster, separated by a
rest interval (frames $601$--$718$).}
\label{tab:phase_durations}
\begin{tabular}{ccccc}
\toprule
Trial & Backswing range & Backswing dur. & Downswing range & Downswing dur. \\
      & (frames)        & (frames)       & (frames)        & (frames)       \\
\midrule
$1$  & $72$--$112$  & $40$ & $112$--$128$ & $16$ \\
$2$  & $128$--$162$ & $34$ & $162$--$179$ & $17$ \\
$3$  & $179$--$212$ & $33$ & $212$--$227$ & $15$ \\
$4$  & $227$--$259$ & $32$ & $259$--$276$ & $17$ \\
$5$  & $276$--$306$ & $30$ & $306$--$323$ & $17$ \\
$6$  & $323$--$355$ & $32$ & $355$--$371$ & $16$ \\
$7$  & $371$--$403$ & $32$ & $403$--$420$ & $17$ \\
$8$  & $420$--$452$ & $32$ & $452$--$469$ & $17$ \\
$9$  & $469$--$501$ & $32$ & $501$--$519$ & $18$ \\
$10$ & $519$--$550$ & $31$ & $550$--$567$ & $17$ \\
$11$ & $567$--$600$ & $33$ & $600$--$617$ & $17$ \\
\midrule
\multicolumn{5}{l}{\itshape rest interval: frames $601$--$718$ ($\approx 3.9$\,s)} \\
\midrule
$12$ & $675$--$719$ & $44$ & $719$--$735$ & $16$ \\
$13$ & $735$--$764$ & $29$ & $764$--$781$ & $17$ \\
$14$ & $781$--$810$ & $29$ & $810$--$827$ & $17$ \\
\midrule
Mean $\pm$ SD & --- & $33.1 \pm 4.1$ & --- & $16.6 \pm 0.7$ \\
Range         & --- & $29$--$44$     & --- & $15$--$18$     \\
\bottomrule
\end{tabular}
\end{table}

\subsection{Phase-separated CHPCA on the skeleton}\label{sec:skel_chpca}

\begin{figure}[t]
\centering
\includegraphics[width=0.92\textwidth]{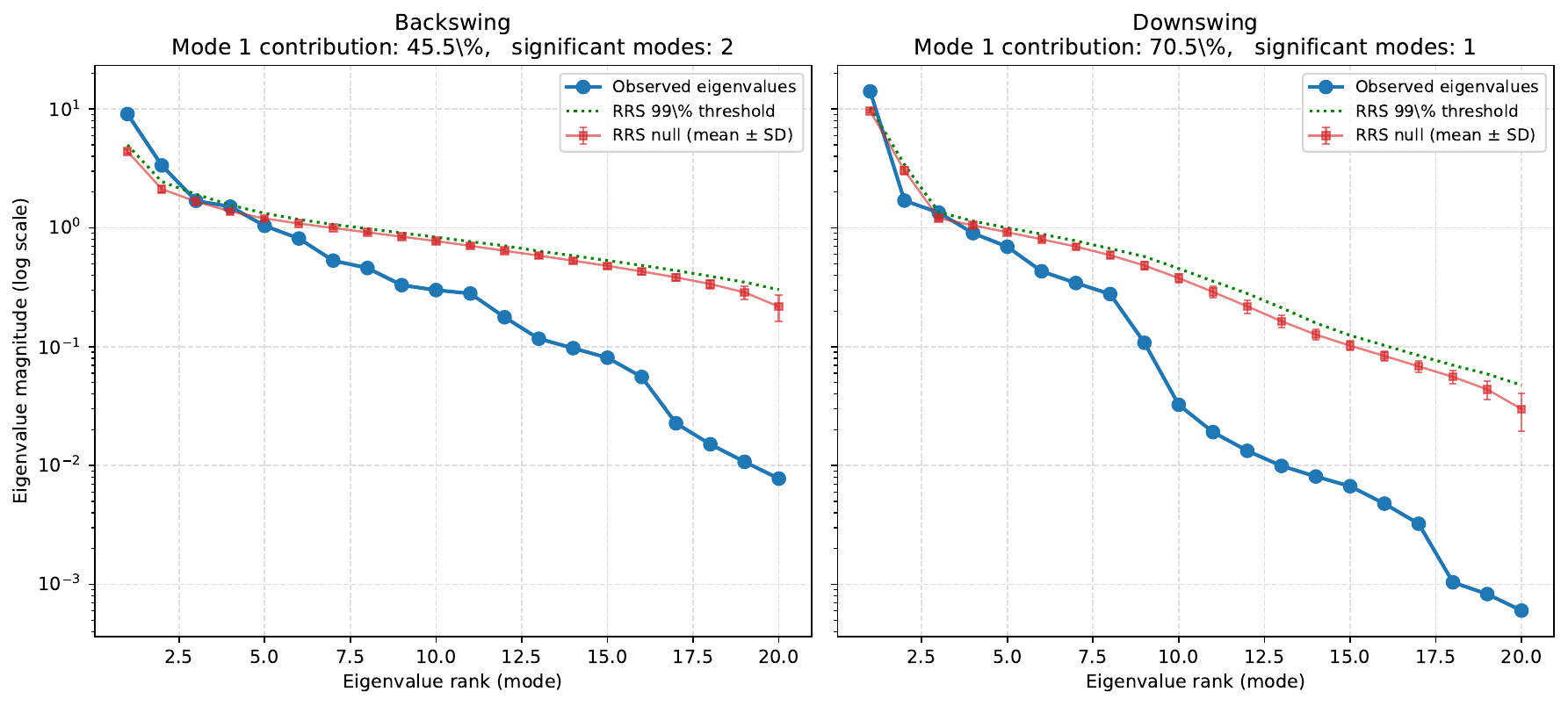}
\caption{Scree plots of the phase-separated CHPCA on the skeleton.  Blue:
observed eigenvalues.  Red: RRS null model (mean $\pm$ SD, $1{,}000$
shuffles).  Green dashed: $99\%$ significance threshold.  In both phases
Mode~1 is far above the null, establishing it as a significant coordination
mode.  Backswing has two significant modes (Mode-1 contribution $45.5\%$);
downswing has one ($70.5\%$).}
\label{fig:scree}
\end{figure}

CHPCA was applied to the $20$-joint skeleton across
$14\ \mathrm{trials}\times 2\ \mathrm{phases} = 28$ segments.  The eigenvalue
spectra and the RRS-based significance assessment are shown in
Figure~\ref{fig:scree}; in both phases the first mode clearly exceeds the
$99\%$ RRS null threshold, identifying Mode~1 as a significant coordination
mode.  The main numerical results are summarised in Table~\ref{tab:skeleton}.

\begin{table}[t]
\centering
\small
\caption{Main results of phase-separated CHPCA (skeleton).}
\label{tab:skeleton}
\begin{tabular}{lcc}
\toprule
Item & A (concatenated) & B (per-trial) \\
\midrule
Mode-1 contribution (backswing)                              & $45.5\%$  & $51.2\%$ \\
Mode-1 contribution (downswing)                              & $70.5\%$  & $74.4\%$ \\
Inter-trial consistency (backswing)\textsuperscript{a}       & ---       & $0.38$   \\
Inter-trial consistency (downswing)\textsuperscript{a}       & ---       & $0.58$   \\
\midrule
\multicolumn{3}{l}{Phase-order reversal (Spearman $\rho$)} \\
$\rho$(backswing, downswing)                                 & $-0.659$  & $-0.654$ \\
$p$ value                                                    & $0.0016$  & $0.0018$ \\
\bottomrule
\end{tabular}

\vspace{1mm}
{\footnotesize \textsuperscript{a} Mean pairwise Spearman rank correlation of
the Hodge-potential vector between trials; closer to $1$ means more
consistent.}
\end{table}

\paragraph{Asymmetry in mode contribution.}
Mode~1 explains about $70\%$ of the total variance in the downswing versus
about $50\%$ in the backswing, suggesting that the downswing is a strongly
coupled motion describable by a single coordination mode, whereas the
backswing is a preparatory and exploratory motion in which multiple
coordination patterns coexist.  Such an asymmetry is not directly measurable
by conventional phase-synchronisation analyses.

\paragraph{Asymmetry in inter-trial consistency.}
The inter-trial consistency is notably higher in the downswing ($0.58$) than
in the backswing ($0.38$).  In the present data (a single subject, $14$
trials), the asymmetry expected from motor-learning theory---that execution
phases are more reproducible and preparatory phases more flexible---is
numerically reproduced in the phase domain.  Generalising this as an
indicator of skill level would require between-subject comparisons; we
therefore treat it here as a case-level observation.

\begin{figure}[t]
\centering
\includegraphics[width=0.92\textwidth]{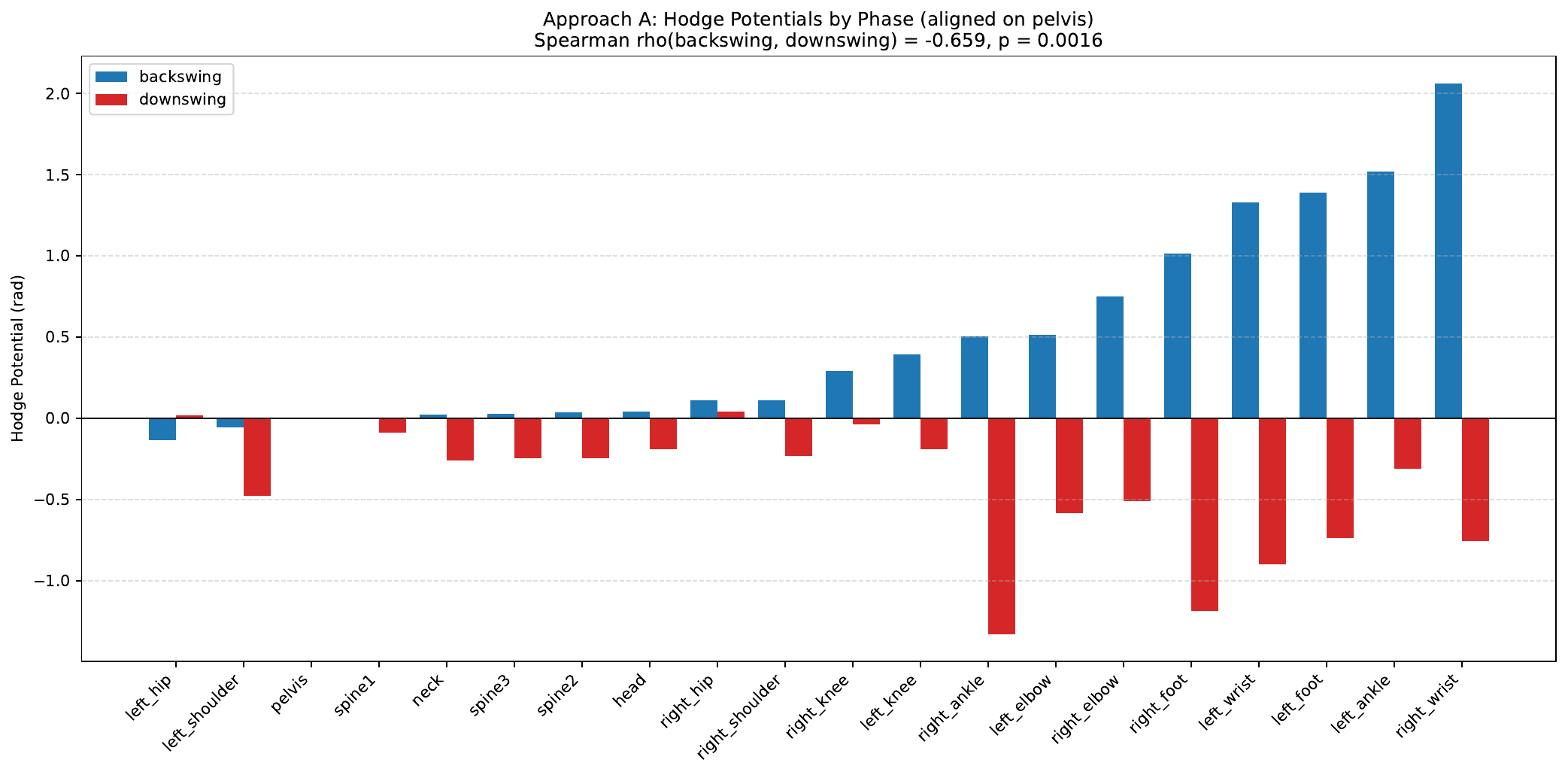}
\caption{Phase-separated Hodge potentials of the $20$ joints (Approach A).
Blue: backswing.  Red: downswing.  The phase sign reverses clearly for most
joints.  The change is particularly large ($\pm 1$--$2\,\mathrm{rad}$) at the
distal segments (wrist, ankle, foot).}
\label{fig:hodge_A}
\end{figure}

\paragraph{Phase-order reversal.}
The Spearman correlation $\rho \approx -0.66$ between the two phases means
that the phase signs reverse for many joints.  This is a physically trivial
behaviour (ordering reverses between ascending and descending motions), but
its faithful reproduction by CHPCA confirms the methodological soundness of
the approach.  The near-identical values across the two approaches
($-0.659$ vs.\ $-0.654$) support the robustness of the result.
Figure~\ref{fig:hodge_A} shows the Hodge potentials of Approach~A.

\subsection{Phase-separated CHPCA on the mesh}

The results of phase-separated CHPCA on all $1{,}079$ SMPL-X vertices are
summarised in Table~\ref{tab:mesh}.  Consistent with the skeleton results,
the Mode-1 contribution is substantially larger in the downswing ($73\%$
vs.\ $43\%$) and the inter-phase order is strongly reversed.  The $p$ value
here is from a Spearman test on rank vectors of length $1{,}079$, which is
bound to be extremely small at this sample size; the effect size
($|\rho| \approx 0.5$) is the substantive quantity.  The vertex-averaged
circular resultant length $R > 0.94$ indicates that the reorganisation
pattern is stably reproduced across trials.

\begin{table}[t]
\centering
\small
\caption{Main results of phase-separated CHPCA (mesh).}
\label{tab:mesh}
\begin{tabular}{lcc}
\toprule
Item & A (concatenated) & B (3 representative trials) \\
\midrule
Mode-1 contribution (backswing)        & $43.3\%$                       & $51.7\%$ \\
Mode-1 contribution (downswing)        & $73.2\%$                       & $77.0\%$ \\
Vertex-averaged phase consistency $R$  & ---                            & $0.95$--$0.96$ \\
\midrule
\multicolumn{3}{l}{Phase-order reversal (Spearman $\rho$)} \\
$\rho$(backswing, downswing)           & $-0.539$                       & $-0.508$ \\
$p$ value                              & $<10^{-10}$                    & $<10^{-10}$ \\
\bottomrule
\end{tabular}
\end{table}

\subsection{Integrated visualisation as networks}

\begin{figure}[t]
\centering
\includegraphics[width=\textwidth]{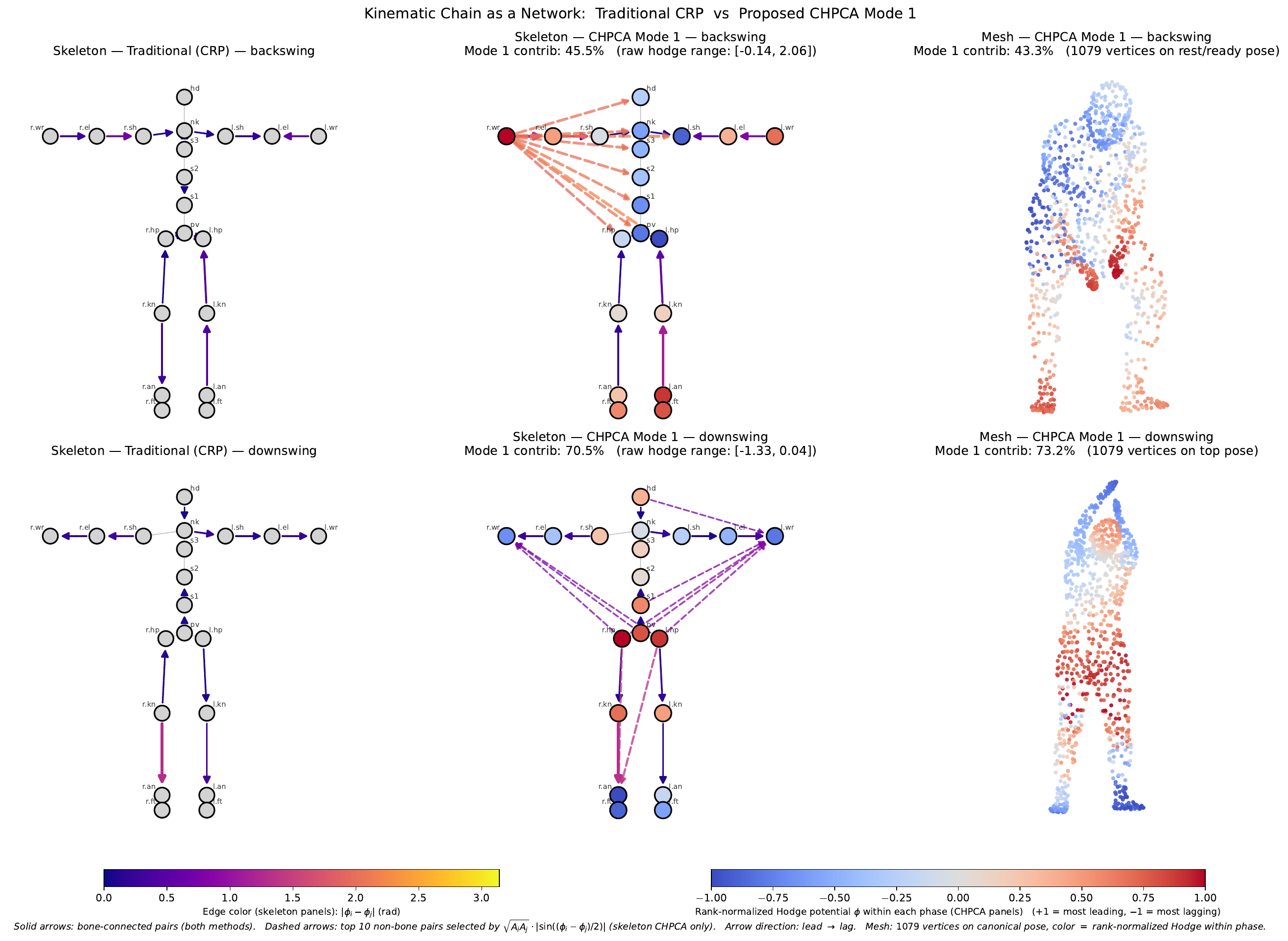}
\caption{Network representations of the kinematic chain.  Left column:
conventional CRP (bone edges only).  Middle column: proposed CHPCA (global
network of the $20$-joint skeleton; solid edges for bone connections, dashed
edges for non-bone pairs with top composite scores).  Right column: proposed
mesh extension (continuous phase field on $1{,}079$ vertices).  Rows:
backswing (top) and downswing (bottom).  Node colour encodes the
within-phase rank-normalised Hodge potential $\phi_i$ (red: most leading,
blue: most lagging).  Edge colour on the skeleton panels encodes
$|\phi_i - \phi_j|$, and arrows point from leading to lagging.  Mesh panels
draw the $1{,}079$ vertices on the pose at the phase onset (backswing: ready
pose; downswing: top pose).}
\label{fig:network}
\end{figure}

Figure~\ref{fig:network} shows a $2 \times 3$ comparison (rows: phases,
columns: methods) of the three resolutions: adjacent CRP, global CHPCA on
the skeleton, and continuous-field CHPCA on the mesh.

\paragraph{Limitations of the conventional method (column 1).}
Because CRP does not define an absolute phase in principle, node colour is
uniform (grey) and the number of edges is limited to the $n-1 = 19$ bone
connections.  Hand dominance, left--right asymmetry and non-adjacent
interactions are not visible.

\paragraph{Global network provided by the proposed method (column 2).}
CHPCA defines an absolute phase, which can be rendered as node colour.
Within-phase rank normalisation immediately reveals the hierarchy.  (i) The
pelvis, spine, neck and shoulders form a collective anchor at
$\mathrm{Hodge}\approx 0$.  (ii) Distal joints (wrist, ankle, foot) show
phase extremes of $\pm 1$--$2\,\mathrm{rad}$, quantifying the Mode-1
structure of ``phase-prominent but low-participation.''  (iii) The
dominant-hand dependence of the motion is quantified in a single global
mode: the dominant (right) side stands out phase-wise in the backswing, while
both sides become more symmetric in the downswing.  The overlaid dashed
arrows visually show that the phase order is not confined to bone adjacency;
however, because the score is structurally larger between anchor and
peripheral joints, individual arrows should not be interpreted quantitatively
as long-range relationships without further statistical tests.

\paragraph{Extension to a continuous phase field (column 3).}
Extending CHPCA to the body-surface mesh changes the description unit from
discrete nodes to a continuous heatmap, while the methodology is identical.
Because CRP would require $581{,}031$ pairs among the $1{,}079$ vertices and
is thus practically infeasible, this continuous extension is a distinctive
advantage of CHPCA.  In the backswing, the right arm and the right side of
the body are red (leading) while the left side and the lower body are blue,
where the dominant hand's lead of the whole-body phase can be seen in
coordination with posture.  In the downswing, the trunk and the lower body
turn red while both arm ends turn blue, and the wave propagation from trunk
to extremities is observable as a spatial gradient.

\subsection{Consistency with existing methods}\label{sec:crp_chpca}

\begin{figure}[t]
\centering
\includegraphics[width=0.65\textwidth]{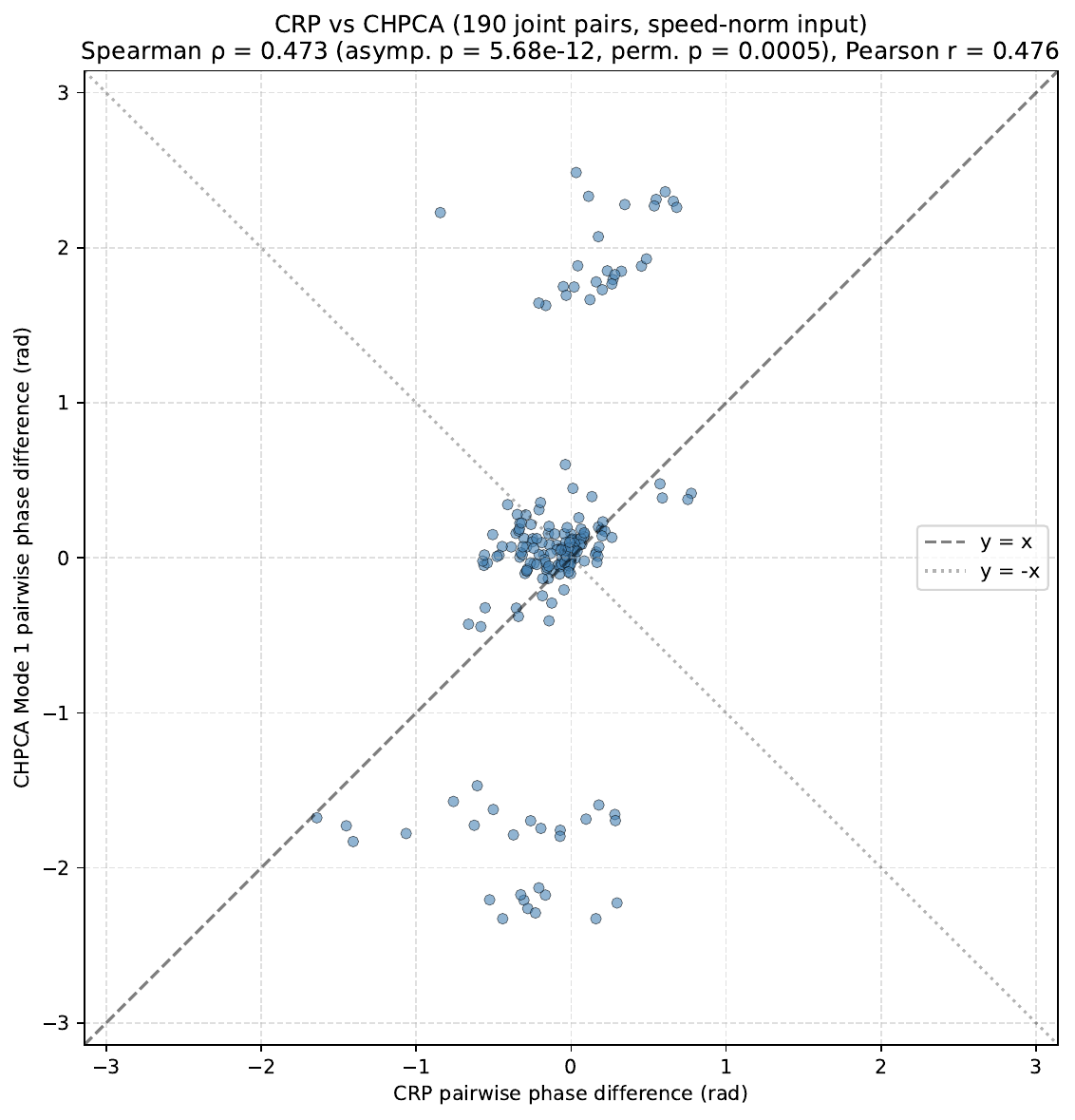}
\caption{Scatter plot of pairwise phase differences between CRP and CHPCA
Mode~1 (all $190$ pairs, full-duration).  Spearman $\rho = 0.473$ (asymptotic
$p = 5.7 \times 10^{-12}$), permutation $p = 0.0005$
($n_\mathrm{perm}=2{,}000$).}
\label{fig:scatter}
\end{figure}

To check whether the pairwise phase relationships of our method are
directionally consistent with conventional CRP, we compared the phase
differences of both methods on all $190$ pairs
(Figure~\ref{fig:scatter}).  To equalise the comparison conditions, both
methods use the $20$-joint speed norm $s_i(t) = \|\bm{v}_i(t)\|$ as input
and compute their pairwise phase differences from the full $900$-frame time
series (without phase splitting).  For CRP, we computed the instantaneous
phase $\varphi_i(t)$ of each joint~\citep{varlet2011-crp-computation} and,
for pair $(i,j)$, wrapped $\varphi_i(t) - \varphi_j(t)$ to $[-\pi,\pi]$ and
took the circular mean over all frames.  For CHPCA, we obtained the Mode-1
complex eigenvector $\bm{u}_1$ from the same input and defined the pairwise
phase difference as $\arg(u_{1,i}) - \arg(u_{1,j})$ (also wrapped to
$[-\pi,\pi]$).  Because both quantities are circular, we converted the
$190$ values into ranks and computed Spearman's correlation; as a robustness
check we also report a permutation $p$ value against the null obtained by
shuffling one side's ranks $2{,}000$ times.

The result, $\rho = 0.473$ with permutation $p = 0.0005$, is significant,
confirming that the two methods agree on the direction of pairwise lead--lag
relationships.  The purpose of this subsection is to confirm that CHPCA at
least preserves the pairwise information available in CRP as a methodological
soundness check; the distinctive value of CHPCA (phase-separated global mode
structure, continuous-field extension) is demonstrated separately in
Section~\ref{sec:energy_amp} and Sections~\ref{sec:skel_chpca}--3.4.  Two
reasons for $\rho$ being away from $1$ are plausible.  First, the two methods
compute similar quantities by different definitions: CRP takes the circular
mean of instantaneous phase differences at each frame, whereas CHPCA
aggregates information into the single mode that explains the most energy
via a covariance structure.  Pair-dependent information carried by modes
other than Mode~1 is not reflected in CHPCA pairwise differences, so
$\rho = 1$ cannot be expected in principle.  Second, because both methods are
computed over the full duration in this comparison, the phase-order reversal
between backswing and downswing (Table~\ref{tab:skeleton},
$\rho \approx -0.66$) is averaged out over the whole sequence, potentially
lowering the attainable $\rho$ in the CRP representation itself.  $\rho < 1$
here can therefore be interpreted not as a primary argument for CHPCA's
uniqueness, but as circumstantial evidence that our method captures
whole-body coordination structure not reducible to a simple pairwise
average.

\subsection{Bridging kinematics and energy mobilisation}\label{sec:energy_amp}

The input to our analysis, the speed norm $s_i(t) = \|\bm{v}_i(t)\|$,
satisfies $KE_i(t)/m_i = s_i(t)^2/2$.  Hence the temporal variance of the
squared speed norm, $\mathrm{Var}(s_i^2)$, is a mass-free proxy for the
variability of kinetic-energy mobilisation at point $i$.  If the CHPCA
Mode-1 amplitude $A_i = |u_{1,i}|$ represents the participation of point $i$
in the Mode-1 coordination pattern, a positive rank correlation between
$A_i$ and $\mathrm{Var}(s_i^2)$ is expected.  In this subsection we verify
this correspondence phase by phase.

\begin{table}[t]
\centering
\small
\caption{Spearman rank correlation $\rho(A_i, \mathrm{Var}(s_i^2))$ between
CHPCA Mode-1 amplitude and the per-point variance of squared speed norm.}
\label{tab:energy_amp}
\begin{tabular}{lcc}
\toprule
Scope                     & backswing                    & downswing \\
\midrule
Skeleton ($20$ joints)    & $-0.080$ ($p=0.74$)          & $+0.708$ ($p=4.7\times10^{-4}$) \\
Mesh ($1{,}079$ vertices) & $-0.048$ ($p=0.11$)          & $+0.719$ ($p<10^{-10}$) \\
\bottomrule
\end{tabular}
\end{table}

\begin{figure}[t]
\centering
\includegraphics[width=0.95\textwidth]{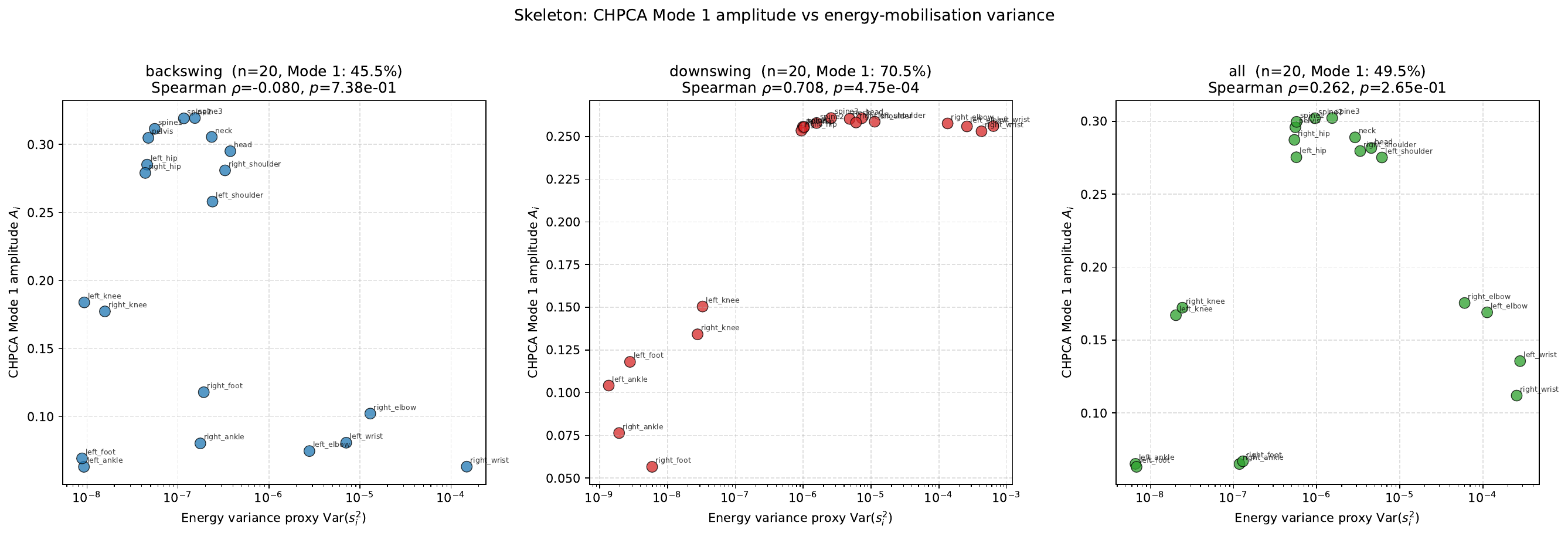}
\caption{Scatter plots of CHPCA Mode-1 amplitude $A_i$ against the
energy-mobilisation variance proxy $\mathrm{Var}(s_i^2)$ on the $20$-joint
skeleton.  Left: backswing.  Middle: downswing.  Right: full duration
pooled.  Horizontal axis is log-scaled.  During the downswing the two
quantities are strongly aligned, whereas during the backswing no
correspondence is observed.  The same asymmetry holds on the $1{,}079$-vertex
mesh (Table~\ref{tab:energy_amp}).}
\label{fig:energy_amp}
\end{figure}

The results show a clear asymmetry consistent across scales
(Table~\ref{tab:energy_amp}, Figure~\ref{fig:energy_amp}).  In the
downswing, both scales yield a positive correlation ($\rho \approx 0.71$):
points with larger Mode-1 amplitude coincide with points showing larger
variability in kinetic-energy mobilisation.  In the backswing, the two
quantities are essentially uncorrelated ($|\rho| \leq 0.08$), so the Mode-1
amplitude captures a phase structure that is independent of energy
mobilisation.  This pattern parallels the asymmetry in Mode-1 contribution
($45\%$ vs.\ $70\%$) and provides energy-based support for the view that the
downswing is compressed into a single low-dimensional mode in both phase and
amplitude.  The near-identical correspondence between the $20$-joint
skeleton and the $1{,}079$-vertex mesh indicates that this asymmetry is
robust with respect to joint definitions and spatial resolution.

The analysis shows that the proposed framework can describe the
spatio-temporal coordination pattern of kinetic-energy mobilisation in the
phase domain from kinematics alone, without the analytical inverse-dynamics
pipeline requiring force plates and inertial parameters.  The method may thus
serve as a means of bridging kinematic and kinetic analyses through
phase structure.

\subsection{Direct comparison between energy phase and Mode-1 phase}\label{sec:energy_phase}

Section~\ref{sec:energy_amp} established an amplitude-level correspondence
between Mode-1 and kinetic-energy mobilisation.  Here we probe the
phase-level correspondence directly.  For each joint $i$, we compute the
kinetic-energy time series $s_i(t)^2$, subtract its mean, and take the
Hilbert transform to obtain an instantaneous phase
$\varphi_i(t)$.  For each pair $(i,j)$ we obtain an ``energy phase
difference''
\begin{equation}
\bar{\varphi}_{ij} = \arg\!\left(
  \frac{1}{T} \sum_t e^{j(\varphi_i(t) - \varphi_j(t))}
\right)
\end{equation}
as the circular mean of instantaneous phase differences across the
concatenated time axis.  The CHPCA Mode-1 phase difference is
$\psi_{ij} = \arg(u_{1,i}) - \arg(u_{1,j})$ (wrapped to $[-\pi,\pi]$).  We
compute Spearman's rank correlation between $\{\bar{\varphi}_{ij}\}$ and
$\{\psi_{ij}\}$ over all $190$ pairs, with a permutation $p$ value against
the null obtained by shuffling one side's ranks $2{,}000$ times.

\begin{table}[t]
\centering
\small
\caption{Pair-level Spearman rank correlation between CHPCA Mode-1 phase
differences and energy-phase differences (skeleton, $n=190$ pairs per
phase).}
\label{tab:energy_phase}
\begin{tabular}{lccc}
\toprule
Phase      & Spearman $\rho$ & Asymptotic $p$           & Permutation $p$ \\
\midrule
Backswing  & $0.831$         & $<10^{-10}$              & $0.0005$        \\
Downswing  & $0.953$         & $<10^{-10}$              & $0.0005$        \\
\bottomrule
\end{tabular}
\end{table}

\begin{figure}[t]
\centering
\includegraphics[width=0.48\textwidth]{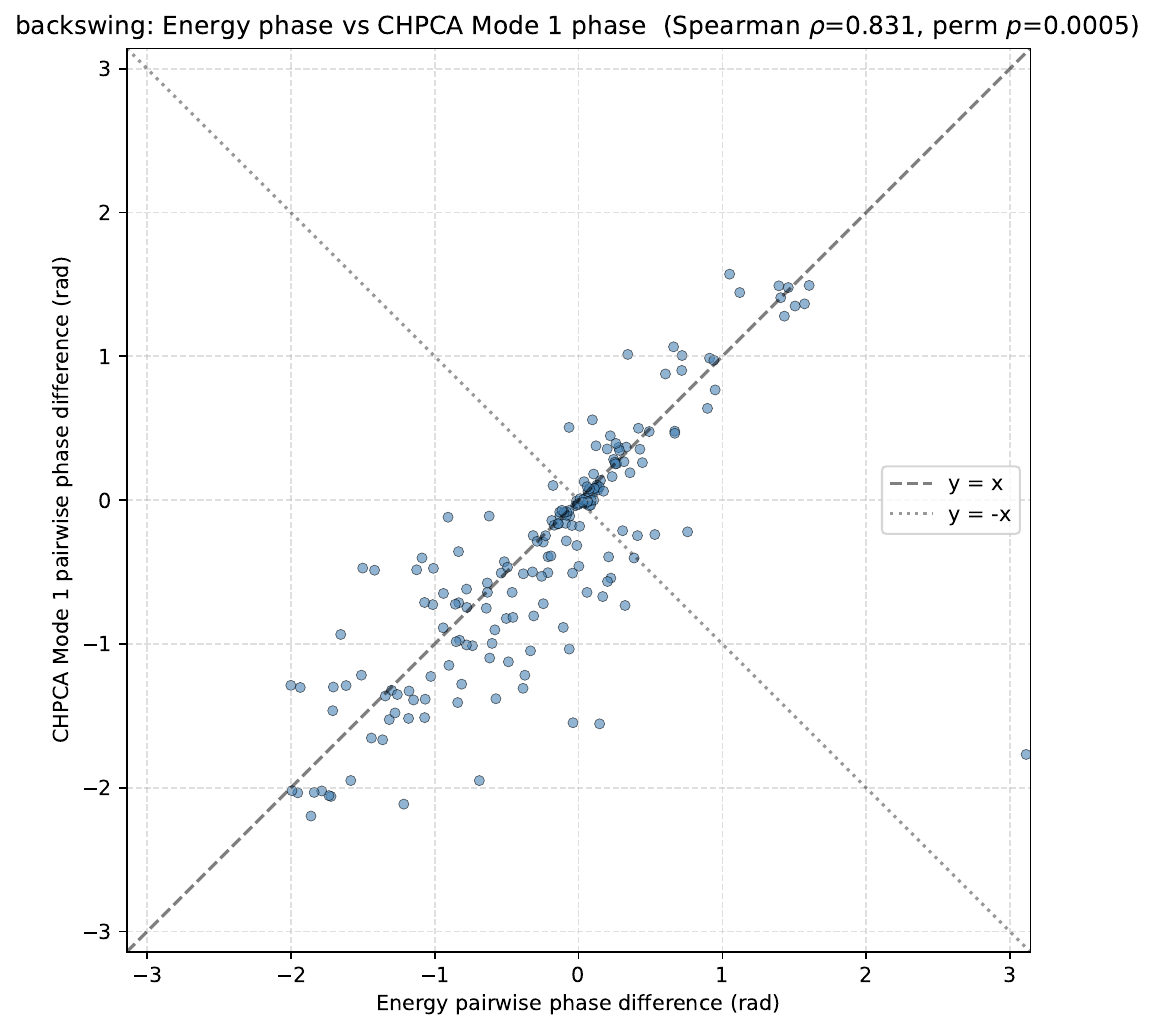}\hfill
\includegraphics[width=0.48\textwidth]{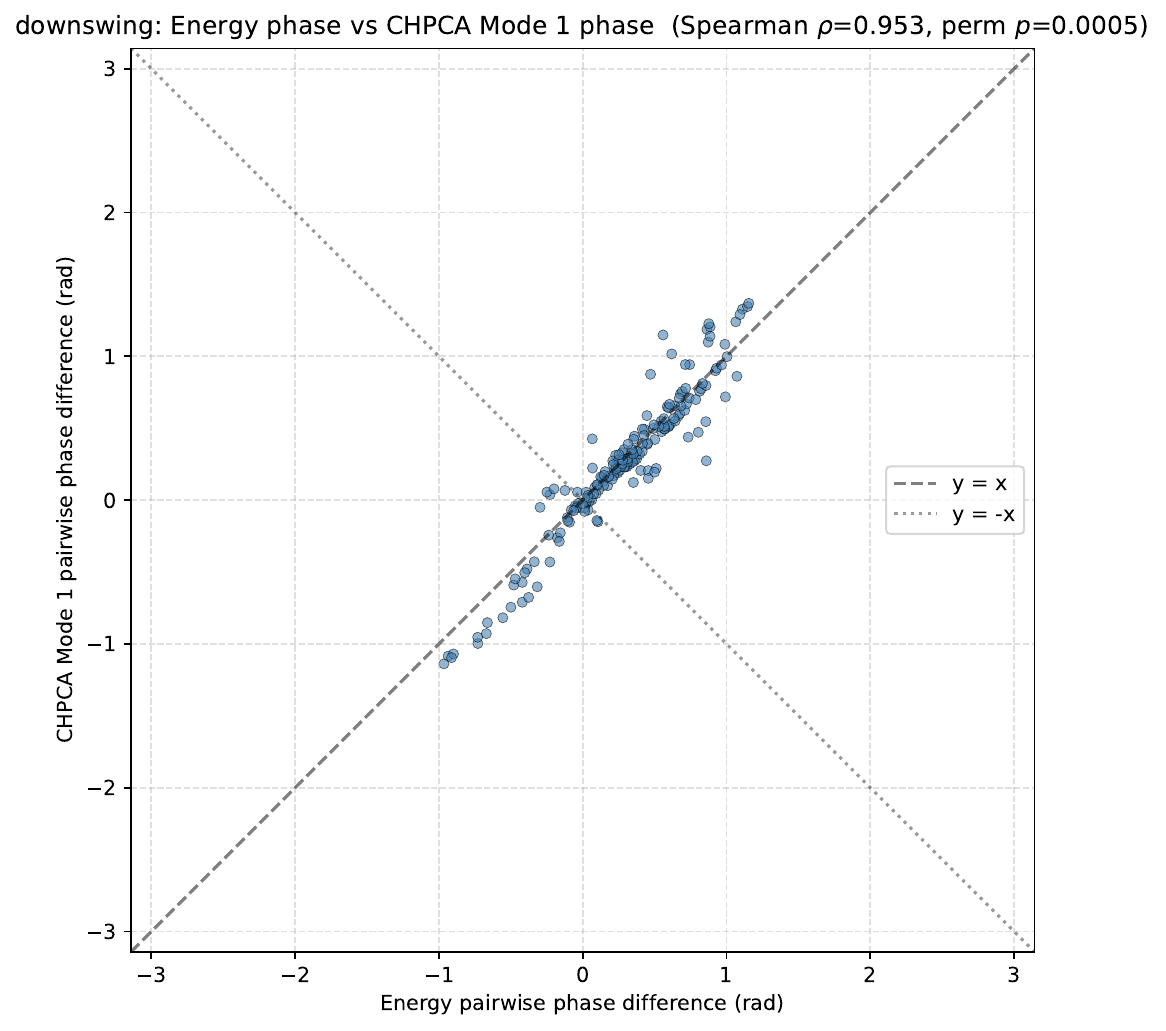}
\caption{Pairwise phase-difference comparison between the Hilbert-transform
energy phase and the CHPCA Mode-1 phase.  Left: backswing ($\rho = 0.831$).
Right: downswing ($\rho = 0.953$).  Each point represents one of the $190$
joint pairs; the $y = x$ diagonal is shown for reference.}
\label{fig:energy_phase}
\end{figure}

The results are striking (Table~\ref{tab:energy_phase},
Figure~\ref{fig:energy_phase}).  In \emph{both} phases the CHPCA Mode-1
pairwise phase differences are very closely aligned with the energy-phase
differences---$\rho = 0.83$ in the backswing and $\rho = 0.95$ in the
downswing.  Combined with Section~\ref{sec:energy_amp}, the following
picture emerges:
\begin{itemize}
\item \textbf{Phase axis.} CHPCA Mode-1 phase structure tracks the timing of
      kinetic-energy mobilisation across joints in both phases.  Even in
      the backswing---where Mode~1 explains only $45\%$ of the
      variance---the Mode-1 phase vector captures the dominant temporal
      order of energy mobilisation.
\item \textbf{Amplitude axis.} CHPCA Mode-1 amplitude tracks the spatial
      pattern of energy mobilisation \emph{selectively} in the downswing
      ($\rho \approx 0.71$) but not in the backswing
      ($|\rho| \leq 0.08$).
\end{itemize}
Thus, the functional asymmetry between the two phases is most accurately
described as: the downswing compresses both the timing (phase) and the
magnitude (amplitude) of energy mobilisation into a single low-dimensional
mode, while the backswing preserves only the timing structure in Mode~1,
with the magnitude distributed across higher modes.  This decomposition of
the coordination structure into ``phase'' and ``amplitude'' axes of the same
complex eigenvector is a distinctive capability of CHPCA that is not
available in real-valued PCA or purely pairwise measures such as CRP.

\section{Discussion}

\subsection{A four-level extension of kinematic-chain description}

Our framework extends the granularity of kinematic-chain description in four
successive levels, each adding information without discarding the previous
level: (1) temporal order of adjacent joints (KS method): pelvis $\to$
thorax $\to$ shoulder $\to$ elbow $\to$ wrist; (2) all-pair phase-difference
matrix (CRP): instantaneous phase differences between any two joints; (3)
global mode as a phase network (this work, Mode~1): the phases of all joints
are jointly represented on a single complex eigenvector, with the magnitude
of the Hodge potential defining node hierarchy and phase differences
defining directed edges; (4) continuous phase field on the body surface
(this work, mesh): discrete nodes are replaced by a continuous body of
$1{,}079$ vertices visualised as the spatial distribution of phase
gradients.  This hierarchy allows qualitative statements such as ``the trunk
is the phase anchor of the whole body,'' ``distal joints show phase
extremes,'' and ``the dominant hand leads in the backswing'' to be
simultaneously tied to numerical indicators on the eigenvector and to
visual evidence.

\subsection{Functional asymmetry in coordination structure}

The asymmetry in Mode-1 contribution ($45\%$ vs.\ $70\%$) and inter-trial
consistency ($0.38$ vs.\ $0.58$) provides, in the present data, a
phase-based indicator that distinguishes a high-dimensional, fluctuating
preparatory motion (backswing) from a low-dimensional, stable execution
motion (downswing).  This asymmetry is an observation that aligns with the
classical notions of the degrees-of-freedom problem (Bernstein) and
equifinality in motor control theory.  However, since the present data come
from a single subject and a single motion type, asserting it as a general
law would require validation through between-subject and cross-motion
comparisons.

\subsection{Bridging kinematics and kinetics}\label{sec:discuss_bridge}

The mechanical-energy and power-transfer
family~\citep{robertson1980-segmental-power,putnam1993-sequential-motion,vaningenschenau1989-biarticular-energy}
has the strength of explicit physical causality, but requires force plates,
anthropometric-table estimates of segmental mass and inertia tensors, and a
rigid-segment model.  These requirements confine the analysis to laboratory
environments and make extension beyond $15$--$20$ rigid segments difficult.
Recent force-plate-free kinetics
estimation~\citep{katsu2024-grf-monocular-video,liu2024-imdy,pbl2025-portable-biomechanics}
and differentiable-physics
biomechanics~\citep{larp2024-neural-physics-pose,cotton2024-differentiable-biomechanics}
relax this constraint gradually but still rely on explicit body models and
do not extend directly to CV-derived dense meshes.

The positive correlation in the downswing between Mode-1 amplitude and the
variance of kinetic-energy mobilisation
(Section~\ref{sec:energy_amp}; $\rho = 0.71$ on the skeleton, $\rho = 0.72$
on the mesh) shows that the spatial pattern of energy mobilisation can be
recovered from kinematics alone as the phase and amplitude of a single
complex eigenvector.  The proposed method thus connects, through phase
structure, laboratory-bound energy analysis with purely kinematic
descriptions obtainable from in-the-wild video.  Because the input is the speed norm, the method cannot recover
absolute energy (Joules) or power (Watts).  Extensions that apply CHPCA
directly to mass-weighted kinetic-energy time series, and multivariate
correspondence analyses with existing energy-based indicators, remain for
future work.

\subsection{Affinity with markerless pose estimation}

The method requires no optical markers and can therefore be applied
directly to practice footage in competitive settings.  The SMPL-X
$1{,}079$-vertex mesh is a dense body-surface representation that is
difficult to obtain with marker-based approaches, and CHPCA processes it
scalably with a single eigendecomposition.  The continuous phase-field
description of the kinematic chain provides an interpretable downstream
analysis for dense estimation outputs from the computer-vision community.

\subsection{Interpretation and mathematical limits of CRP agreement}

Although Section~\ref{sec:crp_chpca} confirmed a significant rank agreement
at $\rho = 0.473$, the fact that $\rho$ does not reach $1$ deserves some
interpretive care.  First, the two methods compute pairwise phase
differences by different definitions: CRP takes the circular mean of
instantaneous phases, whereas CHPCA uses the phase of the single mode that
best explains the covariance structure.  Information carried by
pair-specific components outside of Mode~1 is discarded from CHPCA pairwise
differences, so $\rho = 1$ is not expected in principle.  Second, because
this comparison is computed over the full duration, the phase-order
reversal between phases shown in Table~\ref{tab:skeleton} is averaged away,
potentially compressing the information content of the CRP phase
differences themselves.  Therefore the comparison in this
subsection does not demonstrate CHPCA's unique value; it remains a
confirmation that CHPCA's pairwise information does not conflict
directionally with an established indicator.  A more rigorous evaluation of
correspondence would require multi-level comparison of phase- and
trial-separated pair-difference distributions and the application of
circular--circular correlation measures; this is left for future work.

\subsection{Limitations}\label{sec:limits}

The limitations of the present findings are as follows.
(a) Single subject and single motion type (hammer striking, $14$ trials):
the specific numerical values of Mode-1 contribution and inter-trial
consistency are particular to this instance.  Generalisation to
between-subject differences, skill-level differences, or cross-motion
differences requires comparative validation with data from multiple
participants and multiple motion types.
(b) The phase definition assumes a two-phase periodic motion of ``backswing
and downswing,'' and the automatic segmentation we implemented is a
heuristic built primarily on the striking-side wrist $Z$ coordinate.
Application to other motions such as baseball pitching or a tennis serve
would require motion-specific phase definitions and feature choices.
(c) The analysis takes the speed norm (a scalar) as input, so the $3$D
directional information is discarded at the preprocessing stage.  The
behaviour on motions dominated by direction (e.g., highly rotational
motions) must be verified separately.
(d) Biomechanical interpretations of the mesh phase field (e.g., the
trunk-to-extremity wave propagation) are descriptive observations on a
single subject.

\subsection{Future extensions}\label{sec:future}

(1) Within-session variability in repeated practice: Because the present
method defines statistics over repeated trials, it is well suited to
practice settings in which the same motion is reproduced many
times---baseball batting and pitching, golf swings, and the like.  Mapping
the within-session evolution of Mode-1 contribution and inter-trial
consistency onto changes in fatigue, concentration, or condition would
yield a quantification of fluctuation in the kinematic-chain pattern, and
constitutes the most direct extension of this work toward applied use.
(2) Three-dimensional velocity components: The present analysis discards
directional information by taking the speed norm.  A phase-separated CHPCA
directly on the three-axis velocity components ($v_x, v_y, v_z$), treated
either as $20 \times 3 = 60$ scalar variables or through a suitable
vector-valued extension of CHPCA, would allow distinct axes of coordination
(e.g., vertical vs.\ horizontal) to be compared and is left for future work.
(3) Direct phase comparison with energy time series: Beyond the amplitude
correspondence shown in Section~\ref{sec:energy_amp}, applying the Hilbert
transform directly to the kinetic-energy time series $s_i(t)^2$ and
computing phase differences from those analytic signals would enable a
direct alignment between the CHPCA Mode-1 phase vector and an energy-based
phase vector.
(4) Connection to biomechanical interpretation: The Hodge-potential
distribution obtained as a continuous phase field on the mesh can be aligned
spatially and temporally with muscle activations and joint moments derived
from musculoskeletal models, allowing comparison against anatomical drive
mechanisms.
(5) Multi-subject and multi-motion extension: Whether indicators such as the
Mode-1 contribution asymmetry and inter-trial consistency can distinguish
skill, skill-acquisition progression, and motion style is a natural next
step.  Whether the $0.38$-versus-$0.58$ gap observed here aligns
independently with the invariance hypothesis in motor-learning research is
a concrete empirical question.
(6) Robust phase detection: For applications to real competition footage
without clear periodicity, a hybrid scheme combining coarse motion
segmentation by multimodal large language models with our signal-based
refinement is promising.  An LLM can provide semantic global information
about motion phases (ready, backswing, impact, follow-through), while the
signal side provides frame-accurate boundaries.

\section{Conclusion}

Recent advances in markerless 3D pose estimation, together with the spread
of SMPL-family parametric body models, are shifting the input of sports
motion analysis from a few dozen joints to dense body-surface data of order
$10^3$ vertices.  The standard descriptors of the kinematic chain---
peak-velocity ordering, continuous relative phase, and energy/power
flow---all rely on joint definitions or inertial parameters and do not
directly scale to such dense CV outputs.  Taking the speed norm as the only
required input, we proposed a framework that combines phase-separated CHPCA
with an extension to a continuous phase field on the body-surface mesh, and
applied it to $14$ hammer-striking trials of a single subject.  The main
observations are: (i) the asymmetry of Mode-1 contribution and inter-trial
consistency captures the functional asymmetry between backswing and
downswing on a single mode; (ii) a global phase order not confined to bone
adjacency is jointly represented by a single complex eigenvector;
(iii) the extension to $1{,}079$ body-surface vertices yields a continuous
phase field consistent with the global skeletal structure; and
(iv) a correspondence analysis between Mode-1 amplitude and kinetic-energy
mobilisation variance shows a strong positive correlation in the downswing
($\rho \approx 0.71$ on both skeleton and mesh) and no correlation in the
backswing, positioning the method as a phase-structured bridge between
kinematic and kinetic analyses.  A methodological soundness check against
CRP was demonstrated by a permutation test
($\rho = 0.473$, $p = 0.0005$).  The proposed method extends the description
of the kinematic chain from ``temporal order of adjacent joints'' to
``global phase network'' and ``continuous phase field on the body surface.''
Going forward, we aim to focus on quantifying within-session variability of
coordination patterns (fluctuations driven by fatigue, condition, and the
like) in repeated practice settings, and to develop the framework as an
applied tool along this direction.

\section*{Acknowledgements}
The authors thank the contributors of the Kineo system for making the demo
data publicly available.

\section*{Data and code availability}

The input data used in this study are the publicly distributed Kineo
offline demo data \cite{javerliat2025-kineo}, available from the Kineo
project.  The analysis code (Python) used to generate all figures and
numerical results in this paper is available from the corresponding author
upon reasonable request.

\appendix
\section{Three-axis velocity ablation}\label{app:3axis}

The main analysis uses the scalar speed norm
$s_i(t) = \|\bm{v}_i(t)\|$, which discards directional information.  As an
ablation, we applied phase-separated CHPCA directly to the three-axis
velocity components $(v_{x,i}, v_{y,i}, v_{z,i})$ in the world coordinate
frame.  The input dimension becomes $20 \times 3 = 60$ variables, and the
pelvis $x$-component is chosen as the phase reference.  Table~\ref{tab:3axis}
summarises the comparison against the speed-norm variant.

\begin{table}[t]
\centering
\small
\caption{Speed-norm CHPCA vs three-axis CHPCA (skeleton, phase-separated).
``Corr.'' denotes the per-joint Spearman rank correlation between the
speed-norm Mode-1 phase and the specified quantity from the three-axis
CHPCA.}
\label{tab:3axis}
\begin{tabular}{lcc}
\toprule
Quantity                                         & Backswing              & Downswing             \\
\midrule
Mode-1 contribution (speed norm, $N=20$)         & $45.5\%$               & $70.5\%$              \\
Mode-1 contribution (three-axis, $N=60$)         & $28.7\%$               & $47.4\%$              \\
\midrule
Corr.\ with three-axis $x$-component phase       & $-0.179$ ($p=0.45$)    & $-0.105$ ($p=0.66$)   \\
Corr.\ with three-axis $y$-component phase       & $0.205$ ($p=0.39$)     & $0.284$ ($p=0.22$)    \\
Corr.\ with three-axis $z$-component phase       & $-0.611$ ($p=0.004$)   & $0.311$ ($p=0.18$)    \\
Corr.\ with joint-aggregated three-axis phase    & $0.057$ ($p=0.81$)     & $0.161$ ($p=0.50$)    \\
\midrule
Phase-order reversal (speed norm)                & \multicolumn{2}{c}{$\rho = -0.659$ ($p=0.0016$)} \\
Phase-order reversal (three-axis, joint-agg.)    & \multicolumn{2}{c}{$\rho = +0.675$ ($p=0.0011$)} \\
\bottomrule
\end{tabular}
\end{table}

\begin{figure}[t]
\centering
\includegraphics[width=0.95\textwidth]{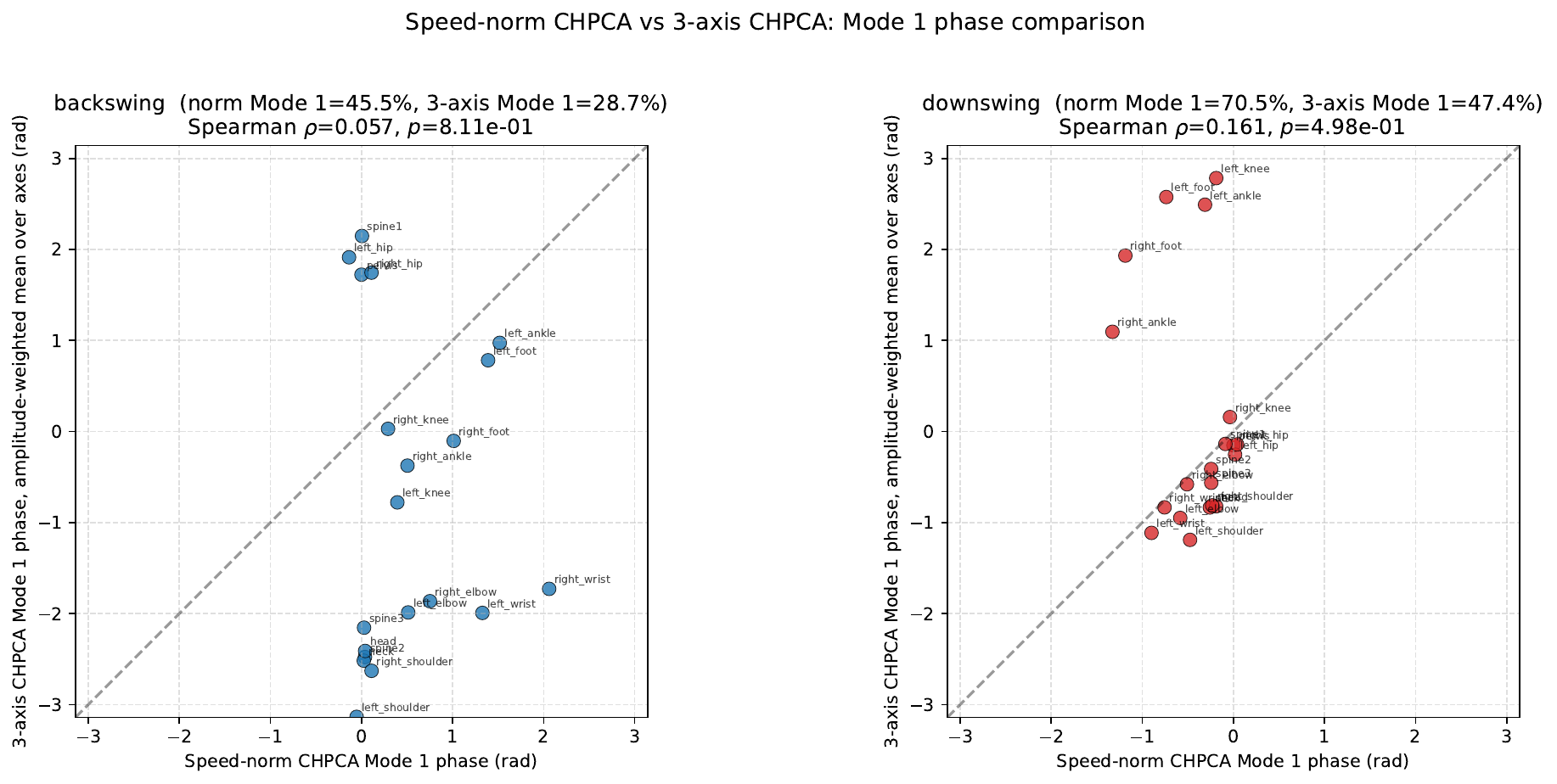}
\caption{Per-joint Mode-1 phase from the speed-norm CHPCA ($x$-axis) versus
the amplitude-weighted circular mean of the three-axis CHPCA Mode-1 phases
($y$-axis).  Blue: backswing.  Red: downswing.}
\label{fig:3axis}
\end{figure}

Three observations follow.
\begin{itemize}
\item The Mode-1 contribution is lower in the three-axis version
      ($28.7\%$ and $47.4\%$) than in the speed-norm version ($45.5\%$ and
      $70.5\%$), as expected from the larger number of variables.  The
      functional asymmetry between backswing and downswing is preserved
      (the downswing Mode-1 share is still substantially larger than the
      backswing share).
\item At the joint level the two representations are only weakly aligned
      (joint-aggregated $|\rho| \leq 0.16$).  This reflects that the speed
      norm compresses three signed axes into a positive scalar, whereas the
      three-axis representation retains directional information.
\item The phase-order relationship between backswing and downswing flips
      sign: $\rho = -0.659$ in the speed-norm version (Table~\ref{tab:skeleton})
      versus $\rho = +0.675$ in the three-axis version.  Physically, this
      occurs because the speed norm is insensitive to the reversal of
      velocity direction between ascending and descending motions, whereas
      the three-axis representation preserves the sign and therefore sees
      the two phases as directionally consistent along the vertical axis.
\end{itemize}

These observations indicate that the speed-norm formulation and the
three-axis formulation describe complementary aspects of the same motion:
the former captures the symmetry of energy mobilisation between phases as
a phase-order reversal, while the latter captures the directional
consistency as a phase-order preservation.  Both views are legitimate and
their difference is informative rather than contradictory.  A
phase-separated CHPCA directly on the three-axis velocity components or a
vector-valued extension of CHPCA is a natural next step
(Section~\ref{sec:future}).

\bibliographystyle{unsrtnat}
\bibliography{references}

\end{document}